# E3C: A Tool for Evaluating Communication and Computation Costs in Authentication and Key Exchange Protocol

Yashar Salami, Vahid Khajehvand [*], Esmaeil Zeinali

Department of Computer and Information Technology Engineering, Qazvin Branch, Islamic Azad University, Qazvin, Iran

**Abstract**: Today, with the development of blockchain and Internet of Things technologies, we need authentication protocols and key exchanges to communicate with these different technologies. Symmetric and asymmetric encryption methods are used to design authentication and key exchange protocols, each of which has different computation costs. In the Internet of Things systems, due to the limited memory and computation power, researchers are looking the lightweight design protocols so that the pressure caused by the computation of protocols can be minimized. Calculating protocols' computational and communication costs was done manually until now, which was associated with human error. In this paper, we proposed an E3C tool that can calculate the computation and communication costs of the authentication and key exchange protocols. E3C provides the ability to compare several protocols in terms of communication and processing costs and present them in separate charts. Comparing the processing and communication costs of classical and modern protocols manually and with the E3C indicate that the E3C can calculate the processing and communication costs of authentication and key exchange protocols with 99.99% accuracy.

## 1. Introduction

Today's world is expanding daily, and this expansion in information technology is increasing [1]. With the spread of the Internet and the advent of new technologies, the Internet of Things has made it possible for physical objects to sense their surroundings and send environmental information to end-users [2]. The Internet of Things has led to the emergence of intelligent vehicles [3] [4], Smart Healthcare [5], UAVs [6], etc., which have played an essential role in human life.

At the same time, the increase in people's need for technology and dependence on it has caused more profiteers to engage in destructive actions. Various attacks on the Internet of Things are expanding, so there is a need to expand the security requirements [7]. Most attacks on computer systems target information confidentiality. Encryption is one of the most critical defense mechanisms against attacks and maintaining the integrity of information. Today, researchers in various fields of security, especially authentication and key exchange, are designing various protocols to improve and secure the Internet of Things environment and use this mechanism. Formal tools must evaluate all authentication and key exchange protocols to validate the protocol.

Formal tools are now widely used by security researchers to evaluate the security of designed

---

[*] Corresponding author. Email: Vahidkhajehvand@gmail.com

protocols. Each protocol is designed after evaluation to assess the resistance to attacks recognized by official tools. The four most commonly used tools are as follows:

AVISPA: Avispa is a tool for automated validation of security-sensitive protocols and applications to attack [8] [9]. The HLPSL language should be used to check the security of protocols in tools Avispa [10]. This language describes protocol scenarios in which each role is independent of the other, receives basic information as input, and can communicate with other roles through defined channels [11].

BAN: BAN is a cognitive logic for analyzing authentication protocols and can model public and private keys [12]. BAN Logic can be used to design cryptographic protocols because a formal language in the protocol design process can prevent breaches [13]. BAN Logic can identify assumptions, results, unnecessary omissions, and what needs to be encrypted in the protocol [14].

Scyther: The Scyther tool, developed by Cas Cremers in 2007, is an automated tool for verifying and forging security protocols [15] [16]. Scyther provides a graphical user interface incorporating the Scyther command-line tool and the Python programming interface [17]. Scyther tool tales Description of the optimal protocol and parameters as input, and output of a summary report and display of a graph for each attack [18].

Proverif: Proverif is a tool for automatically analyzing the security of cryptographic protocols [19] [20]. This tool can prove attacks, which can be used primarily in computer security, and provides the ability to analyze privacy and authentication features [21].

Our studies on tools show that tools Avispa and Scyther, and Proverif are automatic tools. Ban is manual, so coding with this tool is more complex than other tools. In terms of code readability, it can be said that automatic tools have better readability than manual tools. However, none of the tool's understudies can calculate computational and communication costs, so researchers have to do this manually to obtain computational and communication costs. And also, none of the existing tools had the possibility of providing a chart for the obtained results for the users. Table 1 shows a comparison of related work.

Table 1: Comparison Tools.

| Evaluation criteria | Tools Name | | | |
| --- | --- | --- | --- | --- |
| | AVISPA | SCYTHER | PROVERIF | BAN |
| Type | Automatic | Automatic | Automatic | Manual |
| coding | Normal | Normal | Normal | Hard |
| Code readability | Normal | Normal | Normal | Hard |
| Accuracy | Top | Medium | Medium | Low |
| Computing | Not Support | Not Support | Not Support | Not Support |

*1.1 Problem statement*

Sometimes the researcher needs to repeat this protocol several times to compare their protocol with other key exchange and authentication protocols. Manual calculation of computing and communication costs is always accompanied by human error. For this reason, today, we need an efficient tool with any explicit programming language that can calculate the computational and communication costs of authentication and key exchange protocols. The E3C purpose is to reduce manual calculation errors, increase the speed of calculations, and compare the cost of several protocols simultaneously in one chart.

*1.2 Paper Contribution*

- This paper presents an automated E3C tool to calculate the communication and computational costs of authentication and key exchange protocols.

- E3C uses CAS+ language to define authentication and key exchange protocols. E3C allows the user to customize the cost of functions used in authentication and key exchange protocols.

- E3C allows the user to automatically receive the output of their calculations in the form of a chart.

- E3C allows users to compare computational and communication costs of authentication and key exchange protocols.

- To evaluate the E3C, the reference protocols in terms of computational and communication costs have been examined manually with E3C.

*1.3 Paper Organization*

The rest of this paper is organized as follows: section 2 is provided E3C, the Guide Graphical E3C provides section 3, and section 4 includes performance analysis and discussion. Finally, the conclusion is provided in section 5.

**2. E3C**

This section provides information about E3C's architecture, code, and display.

*2.1 Architecture*

The E3C architecture has four parts: coding, calculation, comparison, and display components. Figure 1 shows the E3C architecture.

Coding: The code allows users to implement the proposed protocol using the CAS+ language [22].

Calculation: The computation component calculates the protocol's computing and communication cost based on the implemented code.

Comparison: The comparison component allows users to compare different computational and communication costs of different protocols.

Display: The display component allows the results to be displayed graphically.

*2.1.1 Coding*

In the E3C tool, we needed to use a simple and specific language to define the protocol designed between Alice and Bab to make programming more straightforward and accessible for users, so we used CAS+ language.

One of the features of this E3C tool is that Protocol users can implement their protocol once in the CAS+ language and use it in the Avispa tool for security analysis and in the E3C tool to calculate the computational and communication complexity of the protocol. An example of language code CAS + for protocol Needham

Schroeder Symmetric Key is shown in Figure 2. You can refer to references [22] for more information about CAS +.

As seen in the figure, this language consists of 6 parts, the first part is the name or name of the protocol, and the second part is related to the definition of roles and identifiers used in the protocol. The third part specifies the messages exchanged between the roles. The fourth part specifies the knowledge of each role. In the fifth part, the number of sessions is defined, and the main goals of the protocol are defined in the last part.

> *protocol Needham Schroeder Symmetric Key*
>
> *identifiers*
> A, B, S : user;
> Na, Nb: number;
> Kas, Kbs, Kab: symmetric_key;
> Dec: function;
>
> *messages*
> 1. A -> S      : A, B, Na
> 2. S -> A      : {Na, B, Kab, {Kab, A}Kbs}Kas
> 3. A -> B      : {Kab,A}Kbs
> 4. B -> A      : {Nb}Kab
> 5. A -> B      : {Dec (Nb)} Kab
>
> *knowledge*
> A      : A,B,S,Kas,Dec;
> B      : A,B,S,Kbs,Dec;
> S      : A,B,S,Kas,Kbs,Dec;
>
> *session -instances*
>  [A: alice, B: bob, S: server, Kas:key1, Kbs:key2, Dec: dec];
>
> *goal*
>  secrecy_of Kab [];

*Fig 2. protocol Needham Schroeder Symmetric Key source code.*

*2.1.2 Calculation*

The component written to calculate the cost of computing and communication has been used by default in various articles; the computing computer can be customized by the user.

Communication cost is calculated based on the number of data sent between the communication parties. The execution time of cryptographic functions is considered to calculate the Computation cost, and the total cost is obtained from the sum of the execution time of the used functions.

*2.1.3 Comparison*

Our comparison component used memory management techniques to improve E3C performance, allowing users to compare different protocols with a single click.

After calculating the Computation and communication costs, users can use this component to compare the results in the storage memory and subsequent executions with other protocols in terms of communication and Computation costs. This component minimizes the computational complexity of comparing different protocols.

*2.1.4 Display*

The display component allows users to display the results of calculations in the chart, single and total**.** This component allows users to show the obtained results in a graph with less human error.

*2.2 Workflow*

The primary purpose of E3C is to calculate the composition and communication costs of key exchange and authentication protocols.

The E3C user workflow can be summarized as follows:

1. First, users must implement the protocol using the CAS+ language. If the users already have a

ready protocol, they can open it in the programming environment.

2. Save the protocol with the suffix.CAS+, which ideally specifies the protocol.

3. Users can Set the protocol's symbols and arithmetic mean (ms). Protocol symbols and arithmetic mean can be customized.

4. Run. If the protocol implementation does not have grammatical problems, E3C will display the results of calculations in the output.

5. Users can display the obtained results in the form of charts.

Figure 3 shows the flowchart of workflow E3C.

**3 Guide Graphical E3C**

In this section, the graphical environment of E3C is examined. Figure 4 shows the graphical environment of E3C. The explanation of the different sections is as follows:

1. This section is the toolbar of the tool where the user can perform general operations such as opening, saving, and exiting from the file section. From the execution section, you can perform operations, execute, compare and chart; the help section is provided to the user by the tool provider.

2. This section shows the operations that E3C can support. From this section, users can select the operators they need.

3. This section shows the default symbol. The user can change these symbols according to his interest. Users can define different symbols for Grammar depending on their taste. This section allows users to simplify symbols.

4. This section shows the number of symbols used in the protocol separately for each symbol. After successful execution, it displays all the functions used in the defined protocol by number.

5. This section shows the arithmetic mean for each symbol, which is inferred from paper [23] by default. The user can change the default numbers according to his needs.

6. This section shows the computational results for each symbol after execution. The calculation results of each symbol are obtained by multiplying the number of symbols by the arithmetic mean of each symbol.

7. This is the programming environment for implementing the user protocol. Users can use CAS+ language in this environment to implement their desired protocols to evaluate computational and communication costs.

8. This section of the execution environment is E3C, from which the user can run the protocol or clear the programming environment.

9. This section allows the user to open and save the protocol implemented in the language CAS+.

10. This section allows the user to compare protocols with each other, and also, the user can draw the output results separately in the symbol and total in the chart.

11. This section shows the total results for the user.

12. This section shows the results in the form of a chart for the user.

## 4. Performance analysis

This section evaluates the computational and communication costs of classical protocols such as Wide Mouthed Frog [24] [25], Needham Schroeder Public-key [26] [27], and Otway–Rees [28] [29], and modern SMAK-IOV [30] and CE-SKE [31], and LSKE [32] protocols with the E3C tool and manually. Table 2 shows each function's notations and execution time independently. The execution time of each function is determined based on the reference results [23].

Table 2 shows the notations and execution time of each function.

| No. | Description | Notations | Execution time |
|---|---|---|---|
| 1 | Hash Operation | $T_h$ | $\approx 0.0023$ |
| 2 | Point multiplication | $P_m$ | $\approx 2.226$ |
| 3 | Public-key encryption | $P_e$ | $\approx 3.8500$ |
| 4 | Public-key decryption | $P_d$ | $\approx 3.8500$ |
| 5 | Symmetric-key encryption | $S_e$ | $\approx 0.0046$ |
| 6 | Symmetric-key decryption | $S_d$ | $\approx 0.0046$ |
| 7 | Communication | $>$ | 0 |

### 4.1 Manually

Wide Mouthed Frog Protocol consists of 2 communication stages; this protocol has used the public key three times for encryption and decryption; each public key costs 3.8500, and the total cost is 2se + 2sd *0.0046=0.0184 ms. Needham Schroeder Protocol consists of 3 communication stages; this protocol has used the public key three times for encryption and decryption; each public key costs 3.8500, and the total cost is 3Pe+ 3Pd *3.8500=23.1ms. Otway–Rees Protocol consists of 4 communication stages; this protocol has used symmetric key five times for encryption and decryption; each symmetric key costs 0.0046, and the total cost is five se + 5sd * 0.0046 = 0.046 ms. SMAK-IOV Protocol has used nine public keys for encryption and decryption, the total cost of which is 9Pe+9Pd *3.8500 = 69.3 ms; the communication cost of this protocol for key exchange and authentication is 9. CE-SKE Protocol can perform a key exchange with a communication cost of 3. This protocol uses seven hash functions and six public keys for encryption and decryption. The total cost set is 7Th* 0.0023 + 3Pe + 3Pd*3.8500 =23.116ms. LSKE protocol has a communication cost of 3, and it uses the hash function eight times, 2 Point multiplication, and four times the public key for encryption and decryption for key exchange; the total cost is 8Th *0.0023+ 2 Pm*2.226 + 2Pe + 2Pd*3.8500=19.870 ms. The results of manual calculations are shown in Table 3.

### 4.2 E3C

We are investigating Wide Mouthed Frog, Needham Schroeder, Otway–Rees, modern SMAK-IOV and CE-SKE, and LSKE protocols with the E3C tool. The results of the Wide Mouthed Frog protocol in E3c show that this protocol has a Computation cost of 0.0184 ms and a communication cost of 2. Figure 5 shows the results of the Wide Mouthed Frog protocol. Figure 6 shows the results of the Needham Schroeder protocol. The results of E3c show that the Computation cost is 23.1 ms, and the communication cost is 3. The Otway–Rees protocol calculation results show that the Computation cost is 0.046 ms and the communication cost is 4. Figure 7 shows the results of the Otway–Rees protocol.

Figure 8 shows the results of the SMAK-IOV. The results of E3c show that the Computation cost is 69.3 ms, and the communication cost is 9. The results of the CE-SKE protocol in E3c show that this protocol has a Computation cost of 23.116 ms and a communication cost of 3. Figure 9 shows the results of the CE-SKE protocol. The calculation results of the LSKE protocol show that the Computation cost is 19.870 ms and the communication cost is 3. Figure 10 shows the results of the LSKE protocol.

## 4.3 Discussion

Calculating communication and computation cost in authentication and key exchange protocols is one of the main criteria for the quality of protocols.

The results of manual calculations of the LSKE protocol show that the computational cost is 19.870 ms, and the communication cost is 3. The results of the computational cost are 19.870 ms, and the communication cost is 3 in E3C. In other words, the results of manual calculations are the same as the results of manual calculations and E3C shown in Table 4.

Our observations of the results of manual calculations and Wide Mouthed Frog, Needham Schroeder, Otway–Rees, SMAK-IOV and CE-SKE, and LSKE protocols show that the computation and communication costs are the same. From the same results of manual and E3C calculations, it can be concluded that the accuracy of E3C compared to the manual calculation is close to 99.99%, which shows that E3C can calculate the calculation and communication cost of authentication and key exchange protocols with high accuracy. In manual calculations, the possibility of human error was very high, which was minimized with E3C. E3C allows the user to automatically obtain the output of their calculations in the form of graphs and also allows users to compare the computational and communication costs of authentication and key exchange protocols. Also, this tool allows developers of authentication and key exchange protocols to increase the readability of protocols and reduce human errors by using a CAS+ programming language.

## 5. Conclusion

This paper presents the E3C tool, which is unique for calculating the communication and computational cost of authentication and key exchange protocols. E3C supports the CAS+ language, making implementing different protocols easier. The results obtained from the E3C evaluation show that this tool can calculate the processing and communication costs in authentication and key exchange protocols with 99.99% accuracy, and the calculation speed increases compared to the manual method. In future work, we want to add the ability to adjust the communication Channel Properties and detect Corruption in authentication and key exchange protocols to this tool for E3C development.

Table 3 shows the results of manual calculations.

| No. | Protocol Name | Number of communications | Hash function | Point multiplication | Public key encryption | Public key decryption | Symmetric key encryption | Symmetric key decryption | Total cost | TC (ms) |
|---|---|---|---|---|---|---|---|---|---|---|
| 1 | *Wide Mouthed Frog* | 2 | 0 | 0 | 0 | 0 | *2se* | *2sd* | 2se + 2sd | 0.0184 |
| 2 | *Needham Schroeder* | 3 | 0 | 0 | 3Pe | 3Pd | 0 | 0 | 3Pe+ 3Pd | 23.1 |
| *3* | *Otway–Rees* | 4 | *0* | *0* | *0* | *0* | *5se* | *5sd* | *5 se + 5sd* | *0.046* |
| 4 | SMAK-IOV | 9 | 0 | 0 | 9Pe | 9Pd | 0 | 0 | 9Pe+ 9Pd | 69.3 |
| 5 | CE-SKE | 3 | 7Th | 0 | 3Pe | 3Pd | 0 | 0 | 7Th + 3Pe + 3Pd | 23.116 |
| 6 | LSKE | 3 | 8Th | 2 Pm | 2Pe | 2Pd | 0 | 0 | 8Th + 2 Pm + 2Pe + 2Pd | 19.870 |

Table 4 shows the comparison of the results.

| No | Protocol Name | Manually | | E3C | | Accuracy |
|---|---|---|---|---|---|---|
| | | Computation | Communication | Computation | Communication | |
| 1 | Wide Mouthed Frog | 0.0184 ms | 2 | 0.0184 ms | 2 | 99.99 % |
| 2 | Needham Schroeder | 23.1 ms | 3 | 23.1 ms | 3 | 99.99 % |
| 3 | Otway–Rees | 0.046 ms | 4 | 0.046 ms | 4 | 99.99 % |
| 4 | SMAK-IOV | 69.3 ms | 9 | 69.3 ms | 9 | 99.99 % |
| 5 | CE-SKE | 23.116 ms | 3 | 23.116 ms | 3 | 99.99 % |
| 6 | LSKE | 19.870 ms | 3 | 19.870 ms | 3 | 99.99 % |

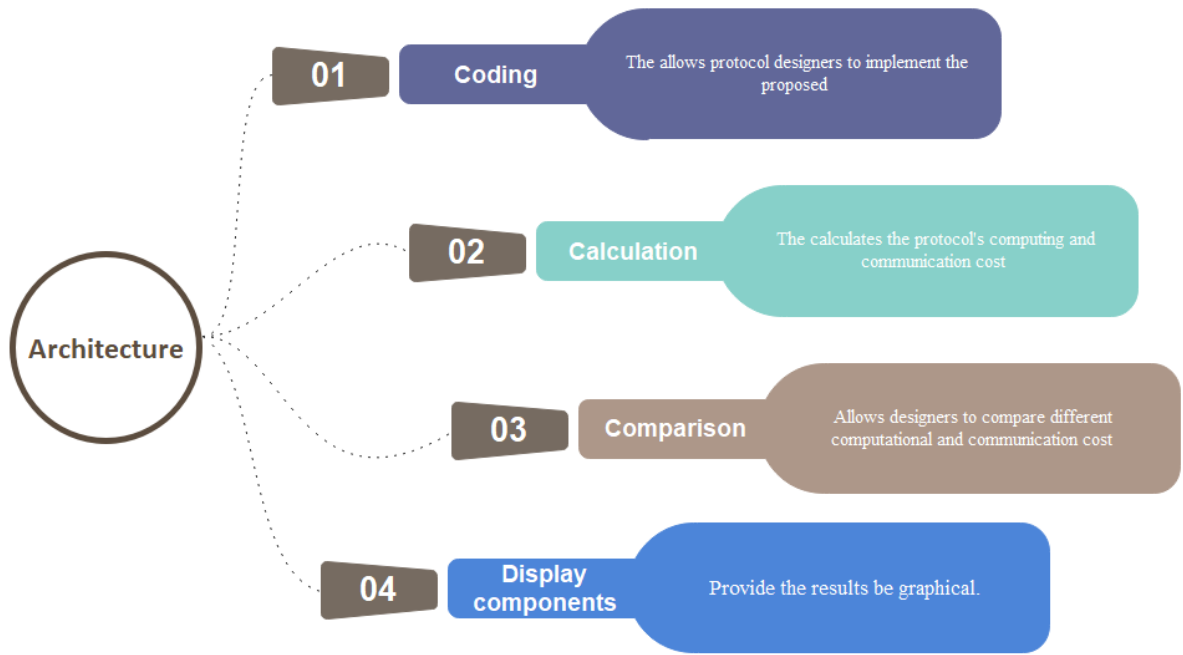

Fig1. E3C architecture.

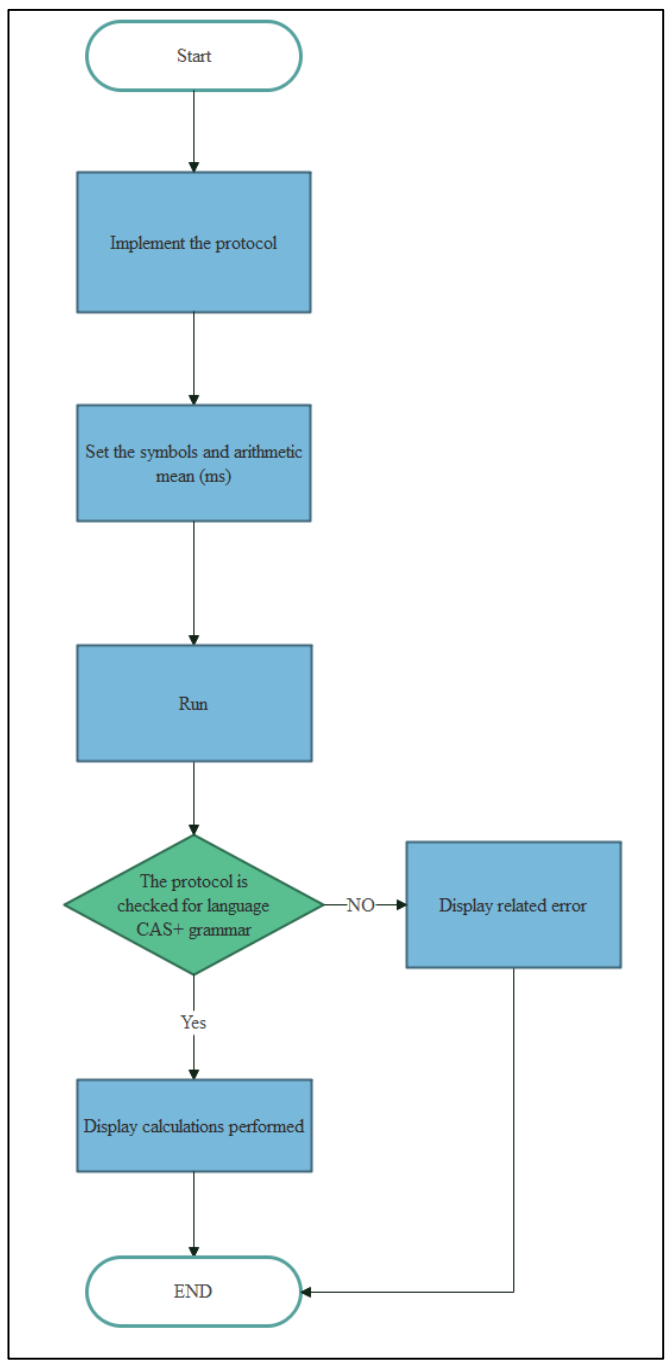

Fig 3. Flowchart of workflow E3C.

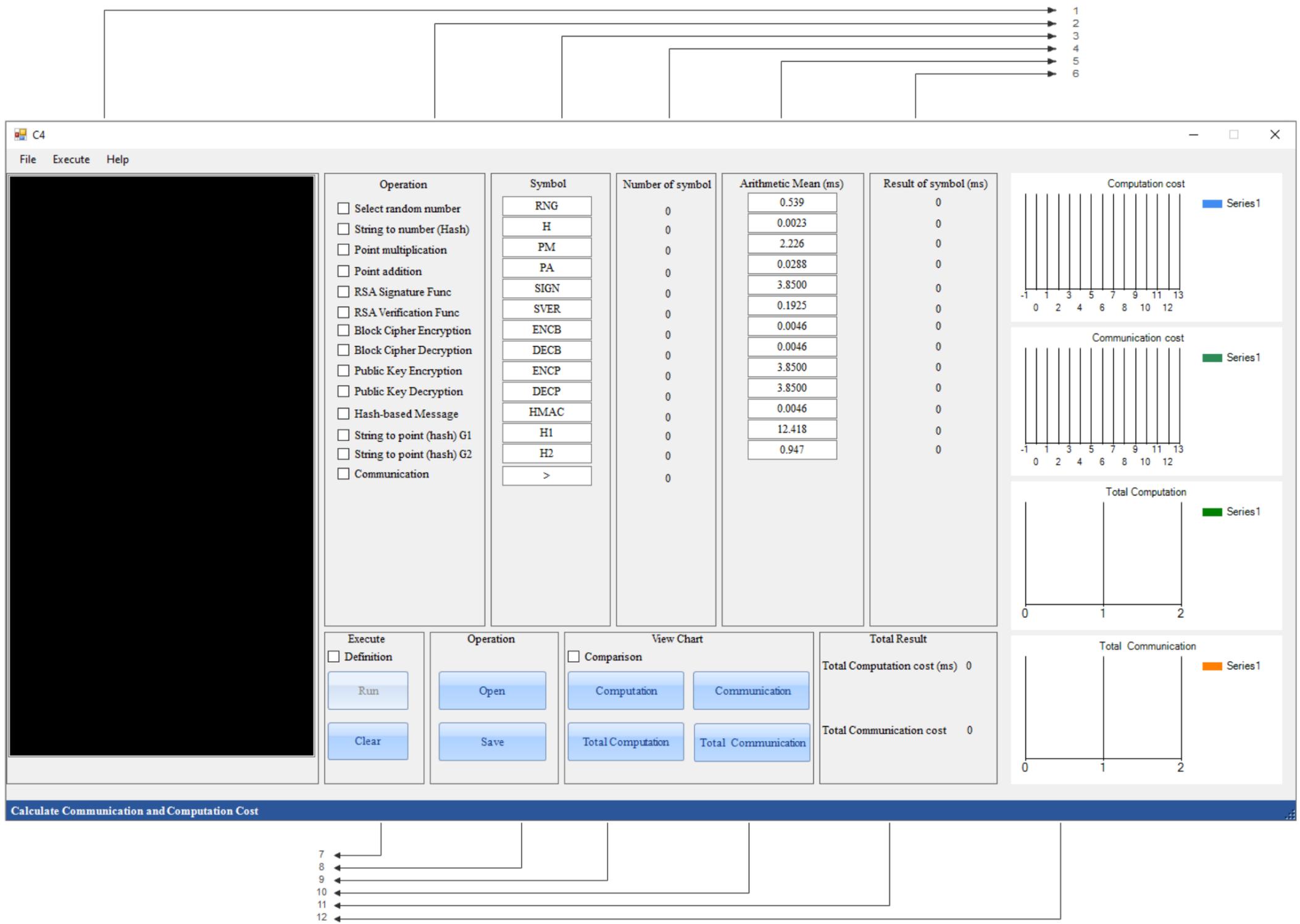

Fig 4. Graphic guide of the environment E3C.

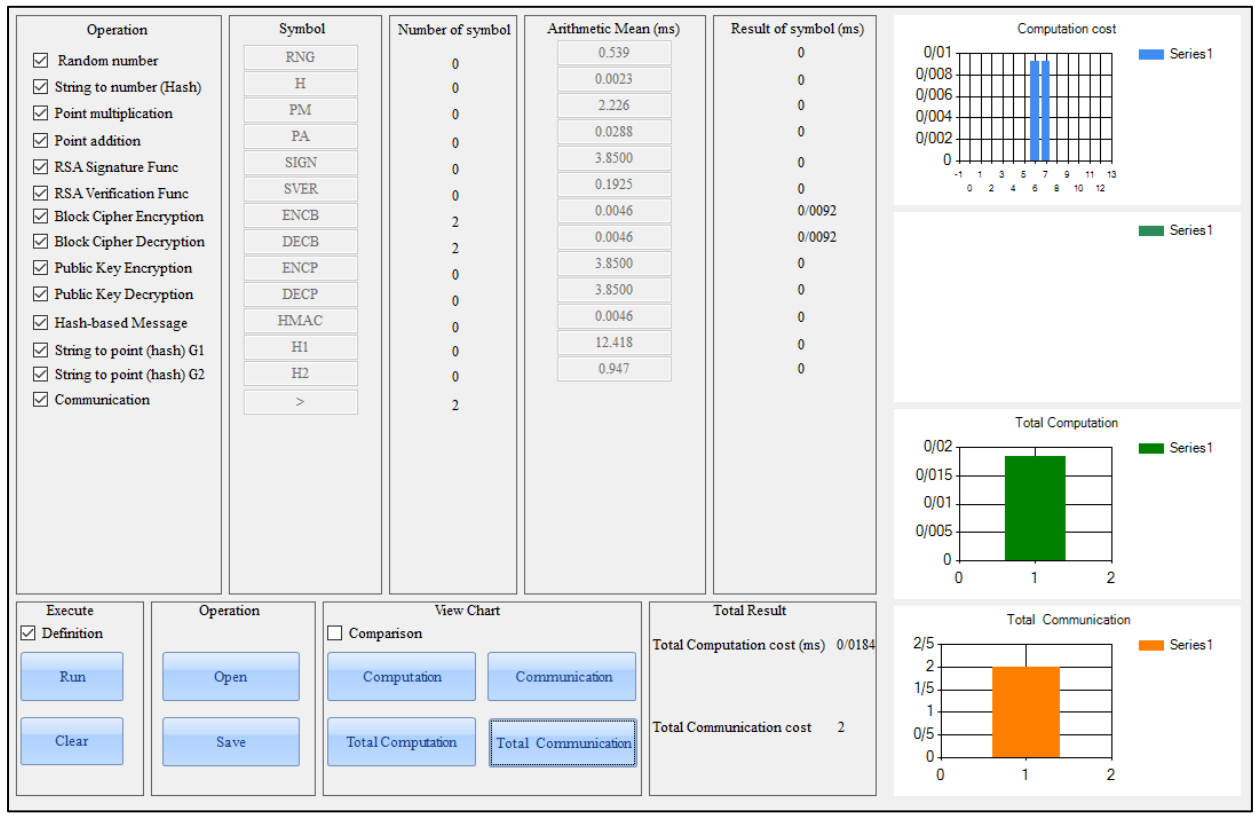

Fig 5. Wide Mouthed Frog protocol evaluation results in E3C.

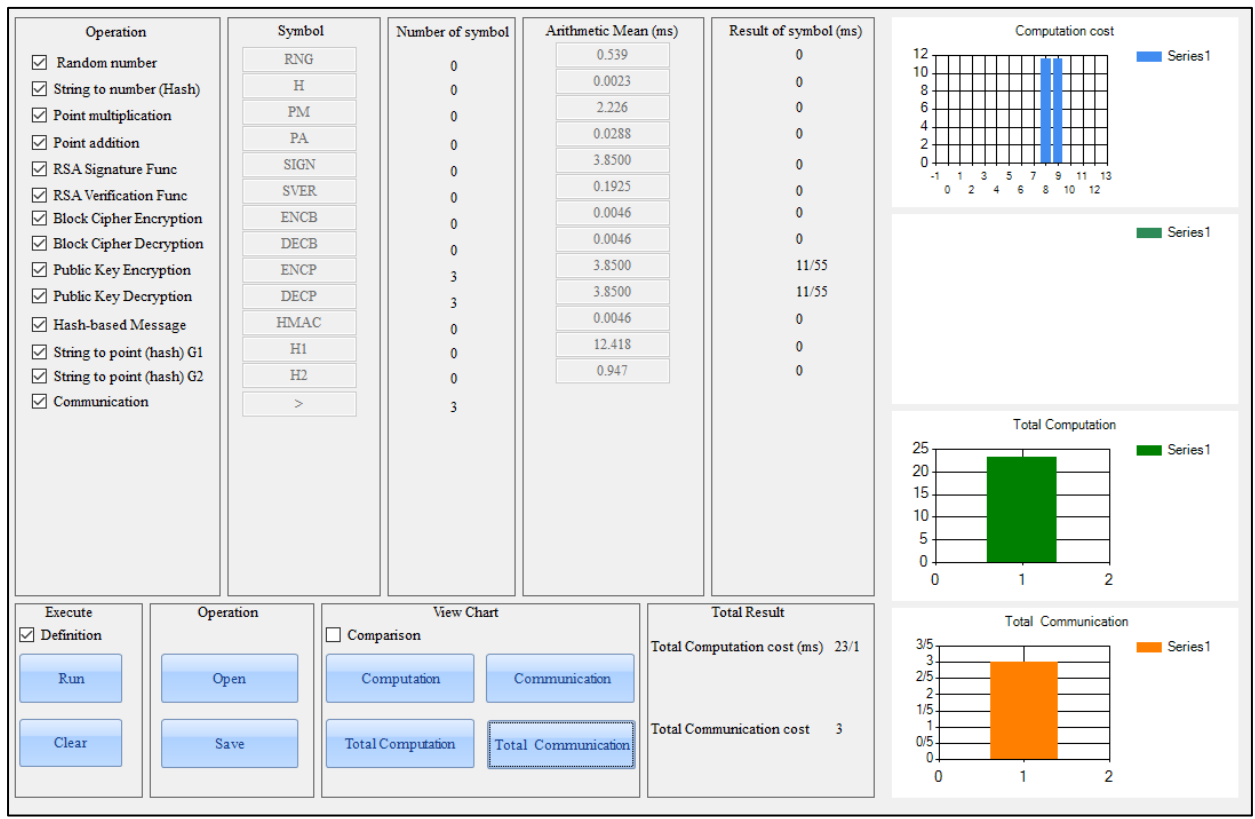

Fig 6. Needham Schroeder protocol evaluation results in E3C.

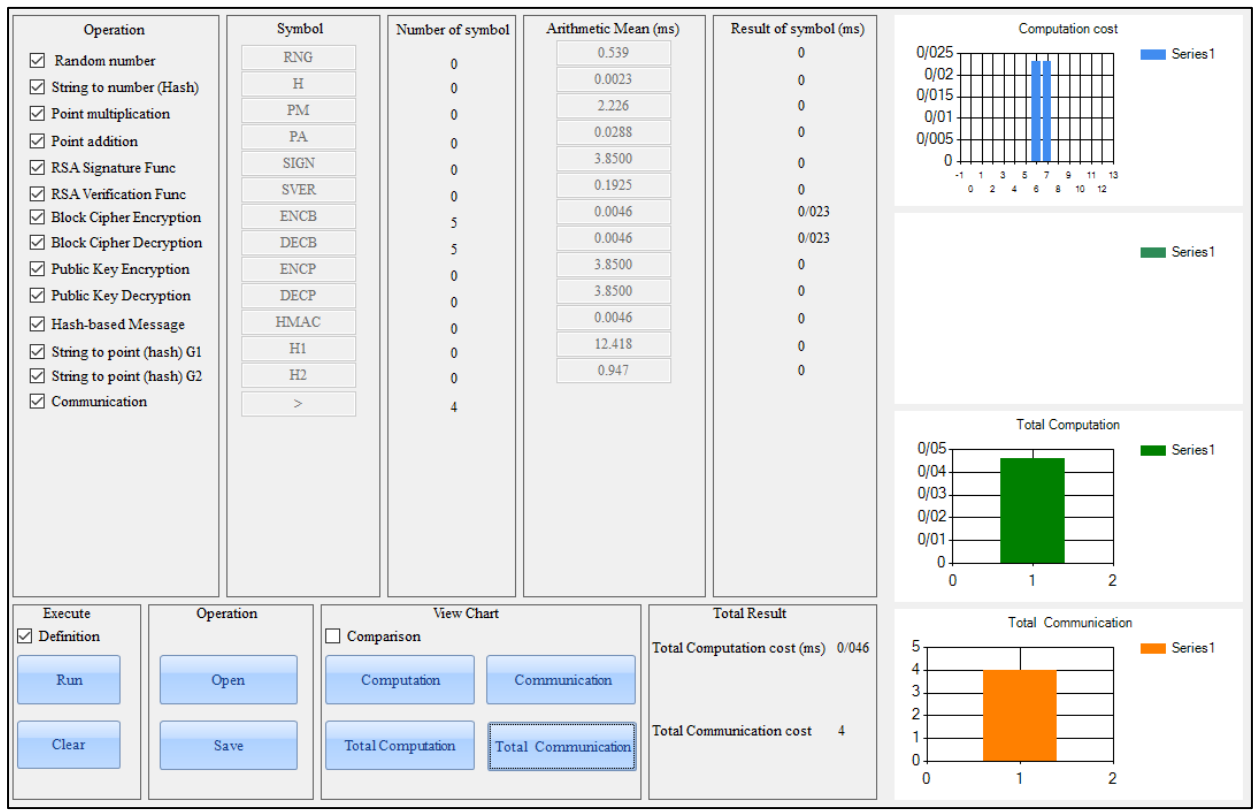

Fig 7. Otway–Rees protocol evaluation results in E3C.

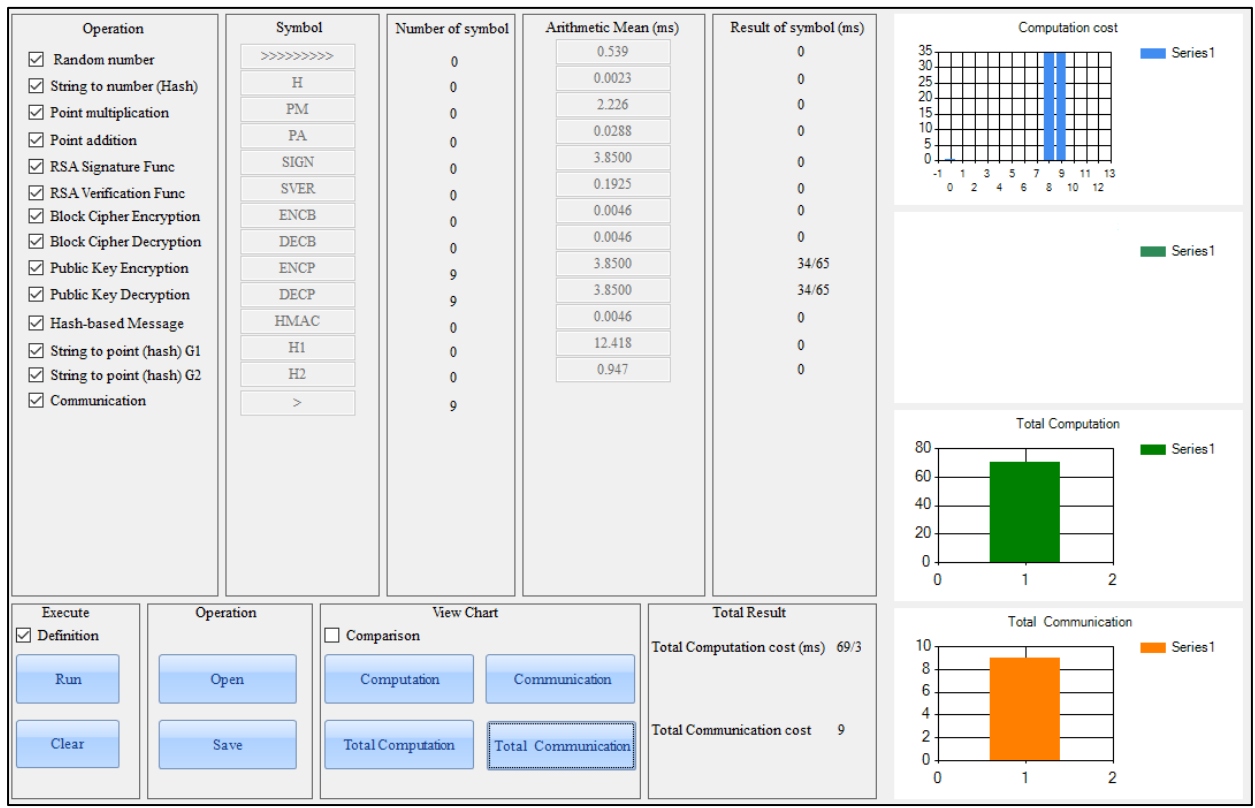

Fig 8. SMAK-IOV protocol evaluation results in E3C.

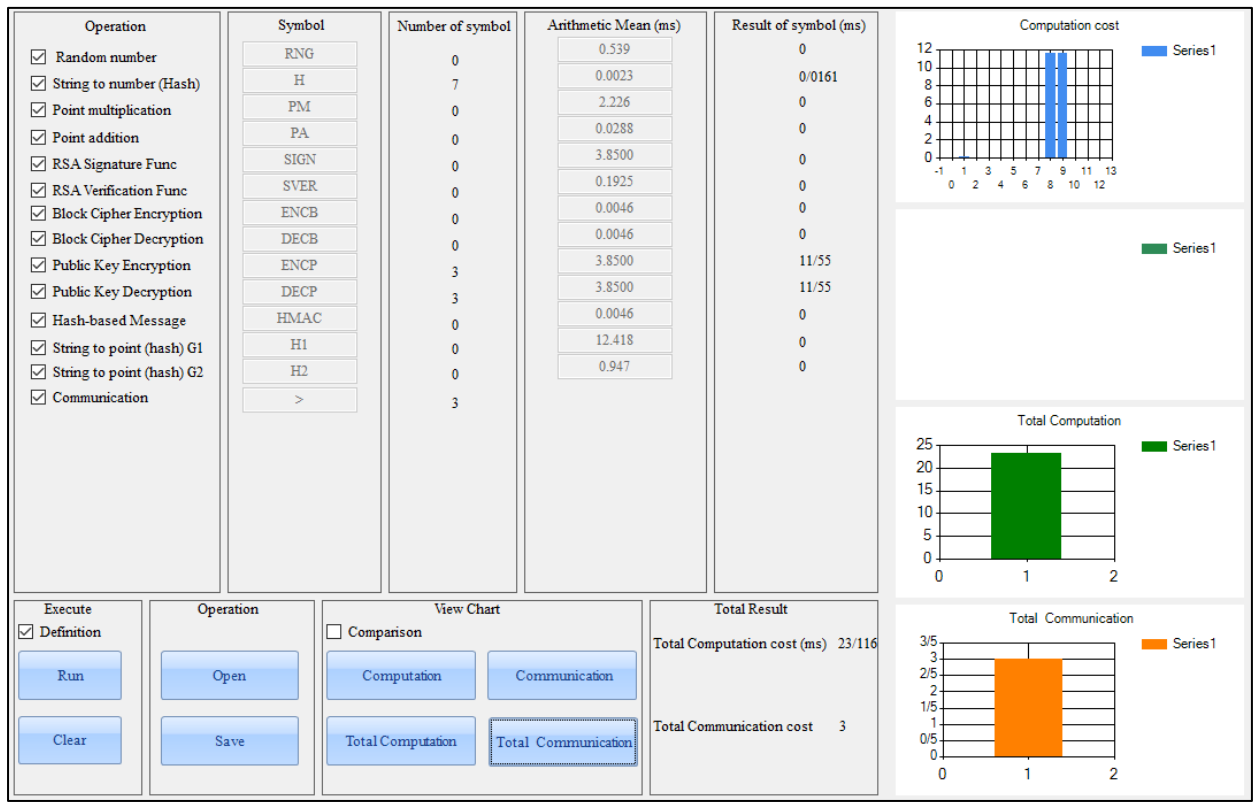

Fig 9. CE-SKE protocol evaluation results in E3C.

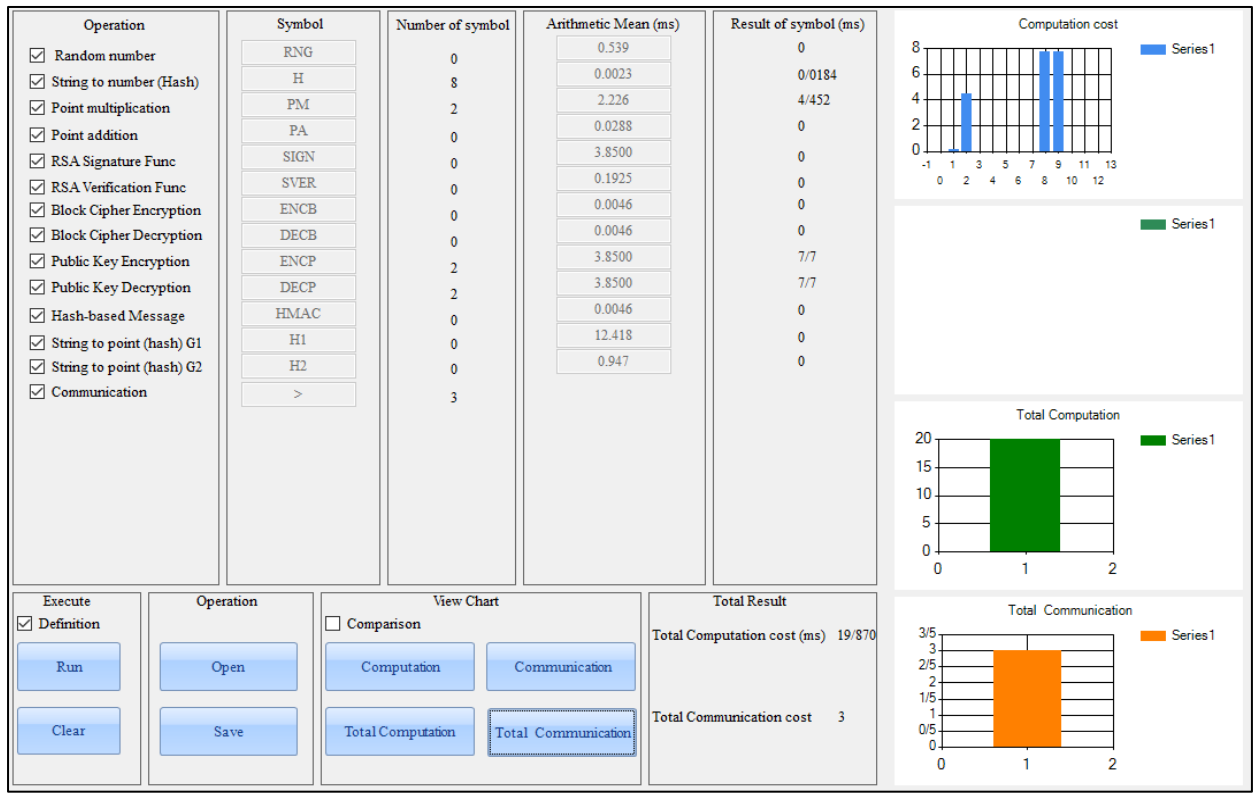

Fig 10. LSKE protocol evaluation results in E3C.